# Protein adsorption onto $Fe_3O_4$ nanoparticles with opposite surface charge and its impact on cell uptake.

M.P. Catalayud[1,2,†], B. Sanz[1], V. Raffa[3], C. Riggio[3], M.R. Ibarra[1,2], G.F. Goya[1,2,†],

1. Nanoscience Institute of Aragón, University of Zaragoza, Spain;
2. Department of Condensed Matter Physics, University of Zaragoza, Zaragoza, Spain.
3. Department of Biology, Università di Pisa, Pisa, Italy.

## *ABSTRACT*

Nanoparticles engineered for biomedical applications are meant to be in contact with protein-rich physiological fluids. These proteins are usually adsorbed onto the nanoparticle's surface, forming a swaddling layer that has been described as a 'protein corona', the nature of which is expected to influence not only the physicochemical properties of the particles but also the internalization into a given cell type. We have investigated the process of protein adsorption onto different magnetic nanoparticles (MNPs) when immersed in cell culture medium, and how these changes affect the cellular uptake. The role of the MNPs surface charge has been assessed by synthesizing two colloids with the same hydrodynamic size and opposite surface charge: magnetite ($Fe_3O_4$) cores of 25-30 nm were *in situ* functionalized with (a) positive polyethyleneimine (PEI-MNPs) and (b) negative poly(acrylic acid) (PAA-MNPs). After few minutes of incubation in cell culture medium the wrapping of the MNPs by protein adsorption resulted in a 5-fold increase of the hydrodynamic size. After 24 h of incubation large MNP-protein aggregates with hydrodynamic sizes of ≈1500 nm (PAA-MNPs) and ≈3000 nm (PEI-MNPs) were observed, each one containing an estimated number of magnetic cores between 450 and 1000. These results are consistent with the formation of large protein-MNPs aggregate units having a 'plum pudding' structure of MNPs embedded into a protein network that results in a negative surface charge, irrespective of the MNP-core charge. In spite of the similar negative ζ-potential for both MNPs within cell culture, we demonstrated that PEI-MNPs are incorporated in much larger amounts than the PAA-MNPs units. Quantitative analysis showed that SH-SY5Y cells can incorporate 100% of the added PEI-MNPs up to ≈100 pg/cell, whereas for PAA-MNPs the uptake was less than 50%. The final cellular distribution showed also notable differences regarding partial attachment to the cell membrane. These results highlight the need to characterize the final properties of MNPs *after* protein adsorption in biological media, and demonstrate the impact of these properties on the internalization mechanisms in neural cells.

---

[†] *Corresponding authors. E-mail: pilarcs@unizar.es ; goya@unizar.es .*

# *INTRODUCTION*

Immediately after magnetic nanoparticles (MNPs) enter a biological fluid, proteins and other biomolecules start binding onto the MNPs surface, leading to the formation of a dynamic 'protein corona' that critically defines the biological identity of the particle.[1] The biophysical properties of such a particle-protein complex often differ significantly from those of the *as formulated* particle, affecting the biological responses as well as the final distribution of the MNPs at the intracellular space. Knowing how physiological medium modifies the final properties of MNPs is therefore decisive for the success of specific applications. Moreover, the lack of knowledge about the new properties can result in unwanted biological side effects. The ability of MNPs to adsorb proteins is expected to depend on the physicochemical characteristics of their surface coating through its affinity for adsorption of ions, proteins and natural organic materials.[2] The proteins adsorbed onto MNP's surface may influence transport across membranes, bringing them into biological entities which they would not normally reach [3, 4], and therefore previous knowledge and quantification of protein-nanoparticle interaction is required for an efficient design of nanoparticles to target cells in vitro. A few theoretical approaches to the dynamics of protein adsorption onto MNPs have rendered some interesting results about the nature of this process,[5] but the simplifying assumptions required for these models to be computationally amenable (e.g. rigid protein structures, single layer formation, etc.)

have so far limited the output when compared to the complexities involved in real experiments.[6]

The interactions of MNPs with cells and tissues are an important factor when considering any potential translation into biomedical applications that require high specificity together with a rapid internalization of the MNPs into the target cells. There is consensus on the fact that surface properties of most MNPs are essential to ensure colloidal stability and that they play a role determining the kind of MNP-cell interactions.[7] But it has only recently been acknowledged that the proteins existent in biological environments can drastically modify the surface of MNPs, therefore deterring the intended therapeutic action. [8]

The surface charge of MNPs is expected to influence the uptake pathway as well as their effective performance.[2] Indeed, it has been demonstrated that, the overall uptake of cerium nanoparticles by human fibroblasts and their respective pathway of internalization depend indirectly on the particle surface charge through the agglomeration resulting of that charge.[9] A series of methodical experiments performed by Safi *et al.*[10] have demonstrated that small γ-$Fe_2O_3$ MNPs can be both adsorbed on the cellular membranes and internalized into human lymphoblastoid cells. These authors tested two types of MNPs coated with citrate ions and poly(acrylic acid) as ligands, but no results on positively-charged MNPs were reported. Due to the 'average' negative charges on the cell surface, MNPs with a positive surface potential are expected to interact in a nonspecific way with binding sites, thus enhancing the efficiency of internalization.[11] [12] [13] [14] However, the cell membrane also presents specific binding sites with cationic receptors that allow interaction with anionic MNPs, in a process described as an "adsorptive endocytosis" pathway.[15] This is in agreement with the well-known uptake of negatively charged MNPs reported by many groups [16, 17] [18] [19].

The above results show that in spite of the large amount of studies on cell uptake of different MNPs and cell types, there is a lack of systematic studies on how surface charge affects the formation of protein corona, and the impact of these changes on cellular uptake. The aim of this work was to perform such a comparative study on the protein adhesion when both positively- and negatively-charged MNPs of similar average size are immersed in protein-rich biological medium. To that end, we performed *in situ* coating of $Fe_3O_4$ MNPs with polyethyleneimine (PEI-MNPs) and poly(acrylic acid)-(PAA-MNPs) by a modified oxidative hydrolysis method, followed by a detailed characterization of their physicochemical properties in the as prepared colloid. The changes of their physical state after incubation in biological medium have been analyzed, as well as these changes on the cellular uptake.

## *Materials and Methods*

**Materials.** All reagents were commercially available and used as received without further purification. Iron (II) sulphate heptahydrate ($FeSO_4 \bullet 7 H_2O$), sodium hydroxide (NaOH), potassium nitrate ($KNO_3$), sulfuric acid ($H_2SO_4$), polyethylenimine (PEI, $M_W$ = 25 kDa) and poly(acrylic acid) (PAA, Mw = 450 kDa) were obtained from Sigma Aldrich.

**Synthesis of PEI- and PAA-MNPs.** The synthesis protocol used for all samples was based on a modified oxidative hydrolysis method, i.e., the precipitation of an iron salt

(FeSO$_4$) in basic media (NaOH) with a mild oxidant. In a typical synthesis, a mixture of 1.364 g of KNO$_3$ and 0.486 g of NaOH was dissolved in 135 ml of distilled water in a three-necked flask bubbled with N$_2$. Then 15 ml of 0.01 M H$_2$SO$_4$ solution containing 0.308 g of FeSO$_4$·7H$_2$O and 0.30 g of polyethyleneimine PEI (25kDa) (previously flowed with N$_2$ for 2 h) was added dropwise under constant stirring. When the precipitation was completed, nitrogen was allowed to pass for another 5 min and the suspension with the black precipitate was held at 90ºC for 24 h under N$_2$. Afterward, the solution was cooled at room temperature with an ice bath, and the solid was separated by magnetic decantation and washed several times with distilled water. For the synthesis of PAA-MNPs, the protocol was the same as described for PEI-MNPs but adding 0.3 g of poly(acrylic acid) PAA (450 kDa) instead of PEI.

**Transmission Electron Microscopy (TEM).** MNPs average size, distribution and morphology as well as SHSY5Y incubated with MNPs were analyzed by transmission electron microscopy (TEM) using a FEI Tecnai T20 microscope and operating at 200 keV. High resolution transmission electron microscopic (HR-TEM) images were obtained by using a FEI Tecnai F30 microscope operated at an acceleration voltage of 300 KV. TEM samples of MNPs were prepared by placing one drop of a dilute suspension of magnetite nanoparticles in water on a carbon-coated copper grid and allowing the solvent to evaporate at room temperature. The average particle size (DTEM) and distribution were evaluated by measuring the largest internal dimension of 200 particles. Cell samples were prepared by treating SHSY5Y cells with PEI-MNP and PAA-MNPs (10 µgml$^{-1}$). After 24 hours of incubation the cells were detached and fixed with 2 % glutaraldehyde solution for 2 h at 4°C and then washed three times in cacodylate buffer (pH 7.2) and treated with potassium ferrocianate 2.5% and osmium tetraoxide 1% for 1 hour at room temperature. After washing, cells were dehydrated with increasing concentrations of acetone 30% (x2), 50% (x2), 70% (x2), 90% (x2) followed by further dehydration with acetone 100%. After drying samples were embedded in a solution (50:50) of EPOXI resin and acetone (100%) overnight, and then for 4-5 hours in resin EPOXI 100%. Sample were dried for 2 days at 60°C and then cut in 70 nm thin slices. STEM-HAADF images were obtained in a FEI Tecnai F30 microscope operated at an acceleration voltage of 300 kV. The microscope was equipped with a HAADF (high angle annular dark field) detector for STEM mode and EDX (X-ray energy disperse spectrometry).

**Zeta Potential.** The zeta potential was evaluated at room temperature on a photo correlation spectrometer (PCS) Brookhaven 90 plus (Zetasizer Nano$^{TM}$ from Malvern Instrument) from a dilute suspension of the sample in water at 0.01 M of KCl.

**Dynamic Light Scattering.** The hydrodynamic diameter distribution of the polymer coated nanoparticles in their aqueous suspensions was obtained using a photo correlation spectrometer (PCS) Brookhaven 90 plus (Zetasizer Nano$^{TM}$ from Malvern Instrument).

**Thermogravimetric Analysis (TGA).** TGA of the powdered samples was performed using TGA/DSC 1 (Mettler Toledo). The analysis was designed at room temperature up to 900 °C fixing a heating rate of 10 °C min$^{-1}$ under a continuous flux of nitrogen. The TG studies of protein adsorption onto PEI and PAA-MNPs were done using the same conditions. These samples were prepared by incubating the nanoparticles with DMEM+15SFB% for 24 h. Then the DMEM+15%SFB was removed by precipitating the nanoparticles with a permanent magnet. The nanoparticles were then dried under air.

**ATR infrared Spectroscopy (ATR-IR).** The ATR-IR spectrum was used to analyze functional groups of Polymers/Fe$_3$O$_4$ nanoparticles and verify their presence on MNPs surface. The spectrum was taken from 4000 to 400 cm$^{-1}$ on a Nicolet Impact 410 spectrometer.

**Magnetic Characterization.** The magnetic measurements were made using a vibrating sample magnetometer (Lake Shore 7400 Series VSM). Magnetization as a function of the field was measured at room temperature up to H = 2 T. Saturation magnetization (Ms) was obtained by extrapolating to infinite field the experimental results obtained in the high range where magnetization linearly increases with 1/H.

**Determination of iron contents in the magnetic colloids.** The $Fe_3O_4$-MNPs concentration in the magnetic colloids was determined by measuring their Fe contents through VIS-UV transmission spectrophotometry (Shimadzu UV-160), based on the thiocyanate complexation reaction: [20]

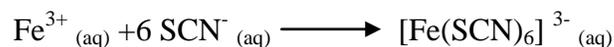

$$Fe^{3+}_{(aq)} + 6\, SCN^-_{(aq)} \longrightarrow [Fe(SCN)_6]^{3-}_{(aq)}$$

PEI-MNPs and PAA-MNPs were dissolved in HCl 6 M-HNO$_3$ (65%) at 50-60 °C during 2 h. Potassium thiocyanate was then added to the $Fe^{3+}$ solution to form the iron-thiocyanate complex, which has strong absorbance at 478 nm wavelength. The iron concentration was determined by comparing the sample absorbance to a calibration curve. As a third independent verification of the iron contents, ICP measurements were done on selected samples. In all cases the values were coincident within error with the amounts inferred from magnetic measurements.

**Cell culture.** Human neuroblast SH-SY5Y cells (ATCC CRL-2266) were cultured in Dulbecco's modified Eagle's medium and Ham's F12 (1:1) with 15% fetal bovine serum, 100 IU/ml penicillin, 100μg/ml streptomycin and 2 mM L-glutamine. Cells were maintained at 37 °C in a saturated humidity atmosphere containing 95% air and 5% $CO_2$. The in-vitro experiments were designed at different concentrations of PEI-MNPs and different incubation. After the incubation time the cells were washed and the modified-DMEM was replaced with ordinary DMEM**.** Control experiments were performed with growth medium without nanoparticles. $CO_2$

**Cell viability assays.** 75x10$^3$ SH-SY5Y cells in exponential growth phase were seeded into a 12 well plate and incubated for 24 h at 37 ˚C with 5% $CO_2$. The media was replaced with increasing magnetic nanoparticle concentrations (0, 5, 10, 20 and 50 μg/mL). The plates where incubated for 24 hr. a) Trypan blue assay: was conducted by diluting 20 $\mu$l of cell samples into trypan blue (1:1). The viable cells were counted. The % cell viability in respect to the control well was calculated whereby the control well was assumed have 100 % viability. b) Flow Cytometry: SH-SY5Y cells of each sample were resuspended in Annexin-binding buffer and stained with 5 $\mu$l of Annexin and 5 $\mu$l of propidium iodide. SH-Y5Y cells were incubated for 15 min in the dark at room temperature. Analysis of the results was performed using a FACS Aria Cytometer and FACS Diva Software.

**Protein Adsorption to MNPs surface.** Adsorption of serum proteins onto the surface of PEI-MNPs and PAA-MNPs was carried out by preparing MNPs (1.5 mg/mL) in DMEM solution containing 15% FBS, to make a total volume of 1 ml**.** The final suspension was sonicated for 30 s to disperse the nanoparticles, and then mixed in a rotating wheel for the different incubation times.

**Quantification of uploaded PEI-MNPs in SH-SY5Y cells.** The amount of MNPs associated to the cells was quantified using a) magnetic measurements, b) UV-VIS absorption spectroscopy and c) inductively coupled plasma mass spectrometry (ICP-MS). For the first method, the amount of magnetic material incorporated per cell was calculated using the saturation magnetization values $M_S$ of the pure colloids (54 and 51 $Am^2kg^{-1}$ for PAA-MNPs and PEI-MNPs, respectively) and the number of cells per sample. SH-SY5Y cells were plated into culture flasks ($1x10^6$cells/flask), at a volume of 5 ml of culture medium. The cells were allowed to adhere for 1 day at 37 °C and 5% of $CO_2$. Then the growth medium was removed and replaced with the medium containing PEI-MNPs. After incubation the cells were washed twice with 2 ml phosphate-buffered saline (PBS), then trypsinized and centrifuged. The precipitate was recovered with 80 µl of PBS and then it was deposited into polycarbonate capsules. The precipitate was lyophilized overnight into the polycarbonate capsule. The magnetic measurements were carried out using a VSM Magnetometer (Lake Shore). Hysteresis loops at room temperature were obtained in applied fields up to 2 T.

To corroborate the values of magnetic material obtained from magnetic measurements parallel experiments were performed with UV-VIS spectrometry, using the same complexation reaction described above, on previously digested pellets with known number of cells, corresponding to those conditions of MNPs concentration and incubation time. As a third independent verification of the iron contents, ICP measurements were done on selected samples. In all cases the values were coincident within error with the amounts inferred from magnetic measurements.

**Dual beam (FIB-SEM) analysis.** To assess the intracellular distribution of MNPs, dual-beam FIB/SEM (Nova 200 NanoLab, FEI Company) analysis images were taken in conditioned samples of SH-SY5Y neuroblasts. SEM images were taken at 5 and 30 kV with a FEG column, and a combined Ga-based 30 kV (10 pA) ion beam was used to cross-sectioning single cells. These investigations were completed by energy-dispersive x-ray spectroscopy (EDX) for chemical analysis.

## Results and Discussion

### Synthesis and characterization of PEI-MNPs and PAA-MNPs.

The magnetic nanoparticles used in this study were synthesized by a modified procedure based on the work of Sugimoto and Matijevic.[21] The method consist of the oxidation of $Fe(OH)_2$ by nitrate in basic aqueous media. We have modified this method by adding the branched polyethyleneimine polymer (PEI, 25 kDa) and Poly(acrylic acid) (PAA, 450 kDa) during synthesis reaction in order to synthesize PEI- and PAA-functionalized $Fe_3O_4$ nanoparticles (labeled hereafter as PEI-MNPs and PAA-MNPs, respectively). The nature of the coating polymer determined the surface charge of the MNPs in the *as prepared* colloids, their resistance to aggregation, and the number of available functional groups on the particle surface. Transmission electron microscopy images (Figures 1a and 1b) showed similar average sizes for both PEI-MNPs and PAA-MNPs. Since the in situ coating determine the morphology of the magnetic $Fe_3O_4$ cores,[22] it is expected that different polymer structures would lead to different particle shape. Accordingly, the PEI-MNPs samples exhibited an octahedral morphology while spherical morphology was observed for PAA-MNPs. The histograms plotted for both samples were obtained after counting a number > 500 of particles. In both cases the histograms could be fitted with a Gaussian distribution, that yielded very similar particle size distributions centered at <d> = 25 nm ($\sigma$=5) and 32 nm ($\sigma$=6) for PEI-MNPs and PAA-MNPs samples, respectively.

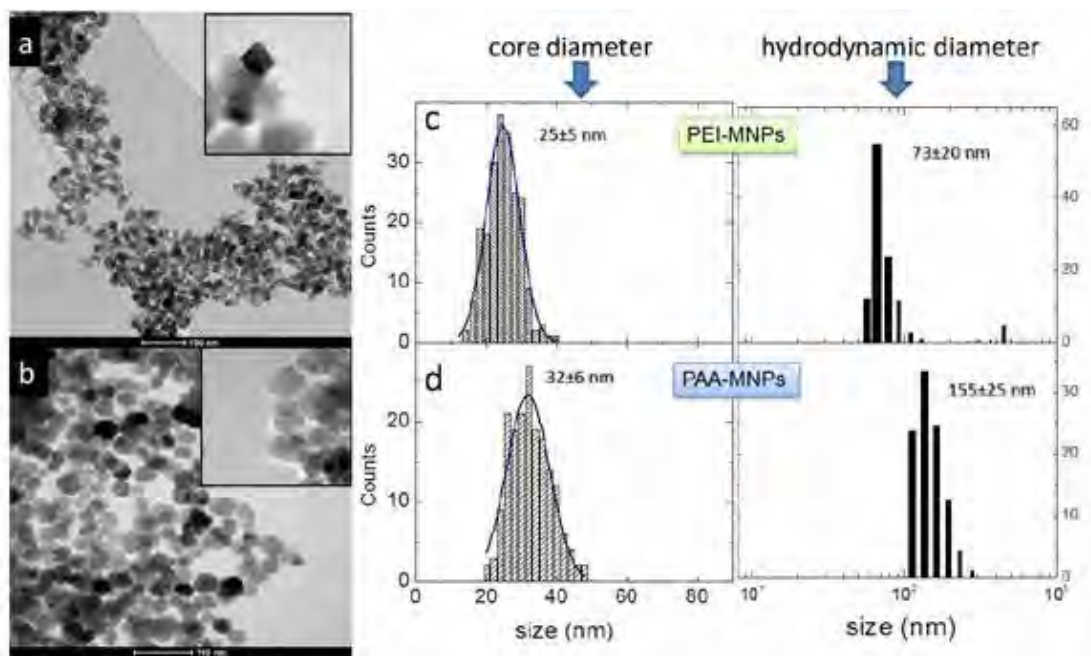

**Figure 1.** *HRTEM images of a) PEI-MNPs and b) PAA-MNPs showing the morphology and overall distribution of particle size. Right panels show the histogram and the fitting Gaussian curves used to extract the average magnetic core size and standard deviation, of c) PEI-MNPs and d) PAA-MNPs samples. The last column displays the hydrodynamic size distributions of the as prepared colloids in aqueous liquid carrier, measured from dynamic light scattering measurements. The data correspond to a number-weighted distribution.*

The dynamic light scattering data obtained in number-weighted distributions (last column of Figure 1), showed that in the *as prepared* water based colloids the degree of agglomeration is higher for PAA-coated MNPs, with average hydrodynamic sizes of 73±20 nm and 155±25 nm for PEI-MNPs and PAA-MNPs, respectively. These values indicate that in aqueous suspension the dispersed units are composed of about 3 to 9 individual particles for PEI-MNPs, whereas for PAA-MNPs this number is larger (~15 to 30 particles). Notably, similar hydrodynamic size values but related to much smaller magnetic cores (6-12 nm) have been reported for PAA-coated magnetic particles, [23] suggesting that it is the polymer that determines the size of the suspended entities. The presence of both PEI- and PAA coating polymers was confirmed by FTIR measurements at room temperature (Figure S1 of supplementary data) by comparing the characteristic IR peaks of the as prepared colloids with those of the pure polymers. The pure PAA spectrum shows the intensity bands of –COOH at 1700 (C=O) and 1212 cm$^{-1}$ (OH) that shifts to 1600 cm$^{-1}$ and disappear respectively for PAA-MNPs. This is due to the PAA attachment onto iron oxide MNPs surface. Similar bands are observed for PEI polymer and PEI-MNPs (REF). The FTIR spectrum of PEI-MNPs, PAA-MNPs and Naked-MNPs exhibit the characteristic bands of the Fe-O bond at 550 cm$^{-1}$.

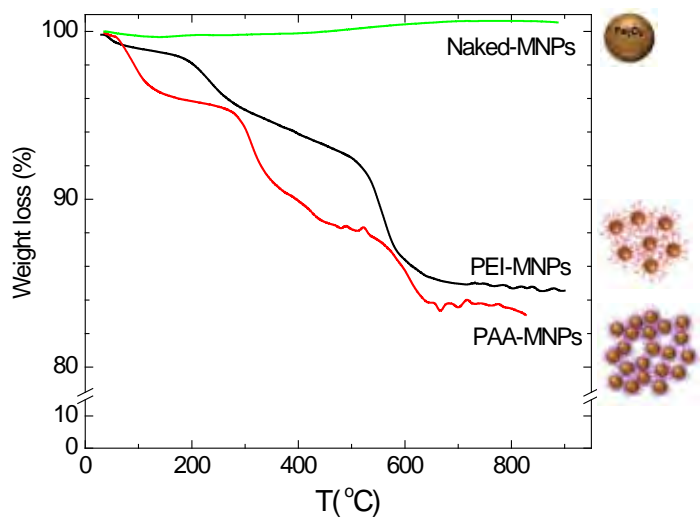

**Figure 2:** *Thermogravimetric analysis of* as prepared *PEI-MNPs and PAA-MNP colloids in water, taken under $N_2$ atmosphere. For comparison, the TGA curve of naked (i.e., without polymer addition) $Fe_3O_4$ cores is shown.*

Thermogravimetric analysis (TGA) performed on the *as prepared* samples of PEI-MNPs and PAA-MNPs colloids showed two regions of maximum rate of mass loss located at 230 and 575 ºC for PEI-MNPs, and at ≈315 and 600 ºC for PAA-MNPs, see Figure 2, yielding a total weight loss of 16% and 17% for PEI- and PAA-MNPs, respectively.

We used a simple model to estimate whether the weight loss of the particles is consistent with a picture of a compact agglomeration of single particles or a less number of MNPs embedded in a polymer matrix. We assumed a single particle of magnetic core of radius $R_1$ nm and a polymer surface layer of thickness t nm, so that the total radius of the particle should be $(R_1+t)$ nm. Analysis of the HRTEM images showed a surface polymer layer of t ≈ 1 nm in both type of MNPs, whereas the average values for the magnetic cores was $R_1$=25 nm and 32 nm for PEI- and PAA-MNPs, respectively. A simple calculation for the expected weight loss, WL, yields

$$WL\% = \frac{100}{1+\left(\frac{\delta_1}{\delta_2}\right)\left[\frac{R_1}{3t}\right]} \qquad \text{Equation I}$$

where $\delta_1$ = 5.17 g/cm$^3$ and $\delta_2$ = 1.08 g/cm$^3$ are the densities of the Fe$_3$O$_4$ cores and the polymer, respectively. The values of the polymer layer thickness t required to obtain the experimentally observed weight losses of 16% and 17% were t = 7.6 and 10.5 nm for PEI- and PAA-MNPs, respectively. These values, together with the hydrodynamic sizes from DLS data of the *as prepared* colloids, are consistent with a picture of agglomerates composed of rather sparsely distributed MNPs within each one, with a total number of N ≈ 12 and 49 magnetic cores per agglomerate in the PEI and PAA-MNPs samples, respectively.

A remarkable stability at room temperature was observed for both colloids along several months, without any noticeable signal of precipitation over time. When compared to the stability of the coatings obtained after through sonication of nude $Fe_3O_4$ MNPs with the same PEI and PAA polymers, the results of in-situ reaction were much better in terms not only of time stability but also regarding resistance to washing procedures with deionized water.

To assess the net particle surface charge, we performed measurements of the ζ-potential of the *as prepared* MNPs in the aqueous medium, as a function of pH (Figure S2 supplementary data). As expected, the positive charge provided by the $NH_2^+$ groups on the PEI-MNPs surface resulted in a high isoelectric point (IEP) value for PEI-MNPs.[22] The carboxyl-rich surface of PAA-MNPs shifted the IEP toward lower pH values.

The hysteresis loops corresponding to PEI-MNPs and PAA-MNPs at room temperature are shown in Figure S3 of the supplementary data. The saturation magnetization was found to be $M_S$ = 51 and 54 $Am^2kg^{-1}$ for PEI-MNPs and PAA-MNPs, respectively. The lower values observed, when compared to those reported for bulk magnetite (90-95 $Am^2kg^{-1}$ at room temperature), it is assigned to a spin disorder effect on the surface of the magnetic nanoparticles resulting in spin canting or misalignment of the local ferromagnetic order. This effect has been reported in several spinel ferrites nanoparticles both in liquid and solid matrix [24, 25], and assigned to broken magnetic superexchange paths mainly at the octahedral 'B' sites of the spinel structure. In the case of polymer-coating as the present MNPs, a similar effect of the polymer layer on the magnetic order has been reported for different organic materials such as oleic acid and explained [26] [27] in terms of covalent bonding of surface Fe atoms to the carbon-based layer that yields the loss of local magnetic ordering.

## *Influence of serum properties on the physico-chemical characteristics of PEI-MNPs and PAA-MNPs*

As mentioned in the introduction, MNPs become coated with proteins and other biomolecules to form a "protein corona" when exposed to a biological fluid. The specific dynamics of the MNP-protein interactions are still not fully understood.[28][29] The overall MNPs-protein corona formation is a multifactorial process that depends on the characteristics of the NPs surface (hydrophobicity, functional groups, etc.) as well as on the interacting proteins and the medium.[30]. Some previous works on the properties of the protein corona have been performed through techniques involving the extraction of the MNPs from the biological medium in which the proteins had attached, and therefore the results are likely to reflect the effect of those proteins covalently-bonded to the MNPs. As noted by Treuel et al.,[31] the situation in biological fluids can be rather different since also those proteins loosely-bounded to the MNP surface contribute to modify their properties and thus a trustable characterization can be done only *in situ*. Therefore, to analyze the evolution of the protein corona formation in biological medium, we have followed the increase of the hydrodynamic size of PEI-MNPs and PAA-MNPs in DMEM+ 15% FBS medium, for increasing incubation times from few minutes to 24 h.

As shown in Figure 3, both types of MNPs have similar hydrodynamic diameter in water (first and second bars on the negative x-axis) with values that indicate a small but measurable degree of agglomeration occurring in the water-based colloid. It is important to notice that the DLS analysis of 'pure' DMEM+15%FBS culture medium, (also included in Figure 3) showed a systematic output of 148±4 nm, originated from the interference of protein structures on the light scattering process. Therefore the size values obtained from colloidal MNPs through this technique should be treated with caution as not only the protein corona but also free-protein clusters are likely to influence the results.

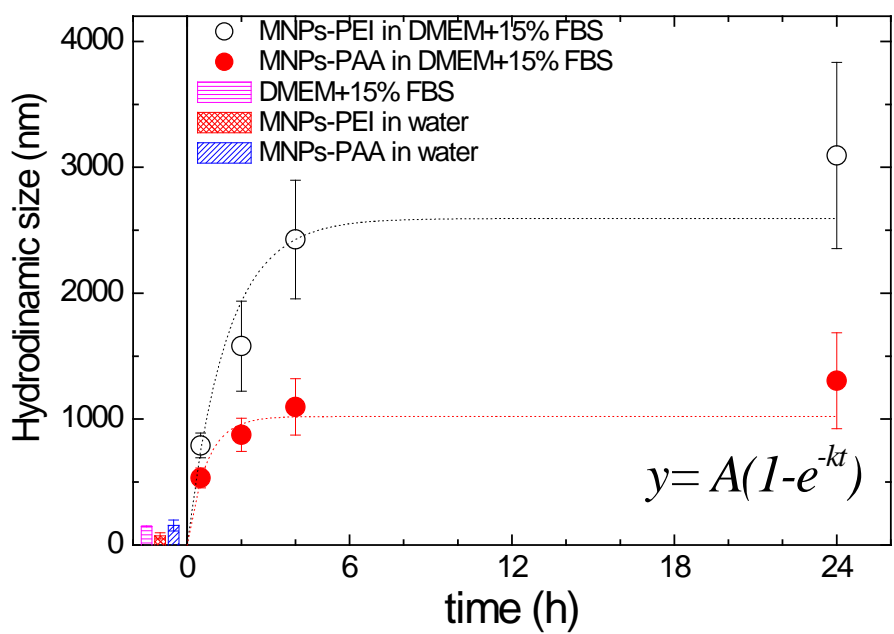

**Figure 3:** *The effect of protein adsorption on the MNPs surface is to enlarge the effective hydrodynamic diameter. Dynamic light scattering measurements showed an increase of size for longer incubation times in biological medium (DMEM+15% FBS). Both PEI-MNPs and PAA-MNPs samples showed the same monotonous increasing profile, with signs of saturation observed at the longest incubation times (24 h).*

Immediately after being dispersed in the culture medium, the hydrodynamic diameter increased from 73±25 nm to ≈900 nm for PEI-MNPs and from 155±44 nm to ≈500 nm for PAA-MNPs. These values increase almost linearly during the first few hours of incubation, and show some signs of saturation after 24 h, when hydrodynamic values up to 1 and 3 µm are observed for PEI-MNPs and PAA-MNPs, respectively. The different agglomeration for both samples should be related to the different electrostatic effect from the negative and positive surface charge of each sample. The PEI-MNPs, having positive $NH_2^+$ groups in the branched structure of PEI polymer, are more affective to bind medium protein and also to crosslink different (negatively charged) units. Consistently, at all incubation times tested the average hydrodynamic diameter of PEI-MNPs sample was larger than the corresponding of the PAA-MNPs. However, electrostatic binding is not likely the only mechanism favoring agglomeration, since PAA-MNPs with negatively charged functional end-groups also showed a considerable degree of agglomeration (up to 1000 nm after 24 h).

Further analysis of the samples in contact with DMEM+15%FBS during 12 h made by TGA was consistent with the process of protein adsorption onto the MNPs. The weight loss curves measured for both MNPs (Figure 4) showed no plateau of constant weight at any temperature.

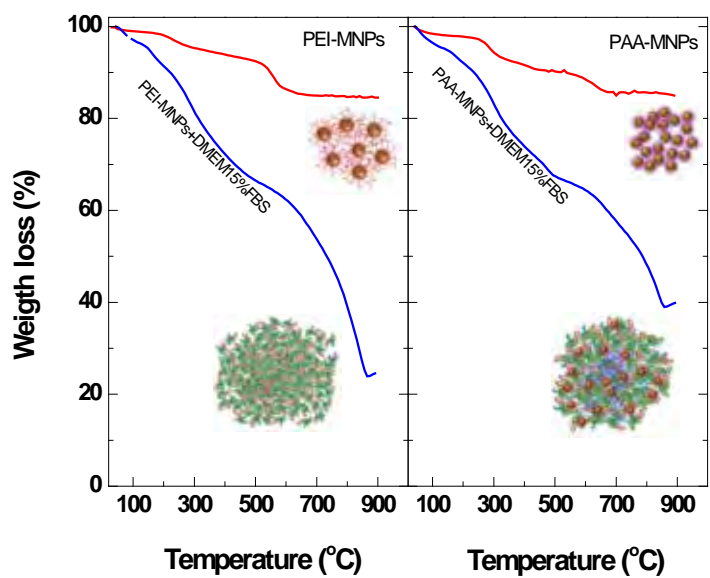

**Figure 4.** *Thermogravimetric data for PEI-MNPs (left panel) and PAA-MNPs (right panel) as prepared and after incubation with DMEM+15%FBS for 12 h.*

Two peaks could be observed in the derivative curves (not shown) indicating that the maximum rate of mass loss occurred at temperatures T = 300-315 ºC and 750-770 ºC. But in contrast with the TGA results of the as prepared colloids in water, the total mass loss of the MNPs after incubation with DMEM+FBS amounted 75% and 60% of the initial mass for PEI-MNPs and PAA-MNPs, respectively. In a similar line of reasoning used for the *as prepared* samples, we considered a 'plum pudding' model of MNPs embedded in a much larger protein-based network to estimate the total size consistent with TGA data. Using the same densities for the $Fe_3O_4$ cores and the coating polymers, and an average value of $\delta_3$=1.006 g/cm$^3$ for the protein network of the DMEM+FBS, the estimated 'thickness' of the protein corona surrounding a single MNPs should be t = 64 and 41 nm for PEI- and PAA-MNPs, respectively. However, the hydrodynamic radii measured by DLS at those incubation times (2 h) were $D_{hyd}$ =

1500 nm (PEI-MNPs) and 895 nm (PAA-MNPs), indicating that these 'aggregates' are actually composed of ≈950 magnetic cores in the case of PEI-MNPs and ≈480 for PAA-MNPs samples. These values of hydrodynamic MNPs size in culture medium, together with its time evolution (see Figure 3) reflects the dynamic nature of the protein adsorption onto MNPs, as illustrated in Figure 5.

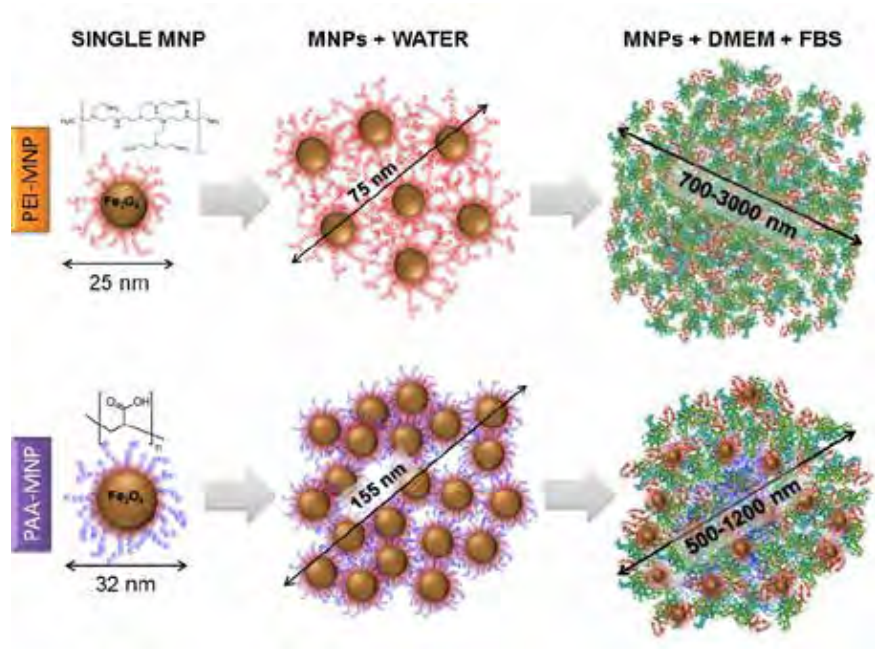

**Figure 5:** *Sketched evolution of the particle agglomeration process for the MNPs when in their as prepared suspension in water (middle column) and when in contact with protein rich medium such as DMEM+FBS (right column).*

As expected, the large amounts of adsorbed proteins ruled the average surface charge during incubation. The ζ-potential evolution of PEI-MNPs and PAA-MNPs (Figure 6) after incubation in DMEM supplemented with FBS showed marked differences. For PEI-MNPs, the value decayed from a positive in water (+30 mV) [22] to negative (-7, -12 mV) in DMEM. In the case of PAA-MNPs the value changes from the as prepared value (-25mV) in water to a less negative value of -11 mV. In order to study the strength of the interaction between the nanoparticles and the proteins, they were incubated in DMEM followed by vigorous washing to remove the unbound proteins. After the washing, it was found that ζ-potential value of PAA-MNPs was approximately -20 mV, similar to the value measured in water. On the contrary, PEI-MNPs still exhibit negatives ζ-potential values, indicating stronger attachment of the proteins onto PEI-MNPs surface.

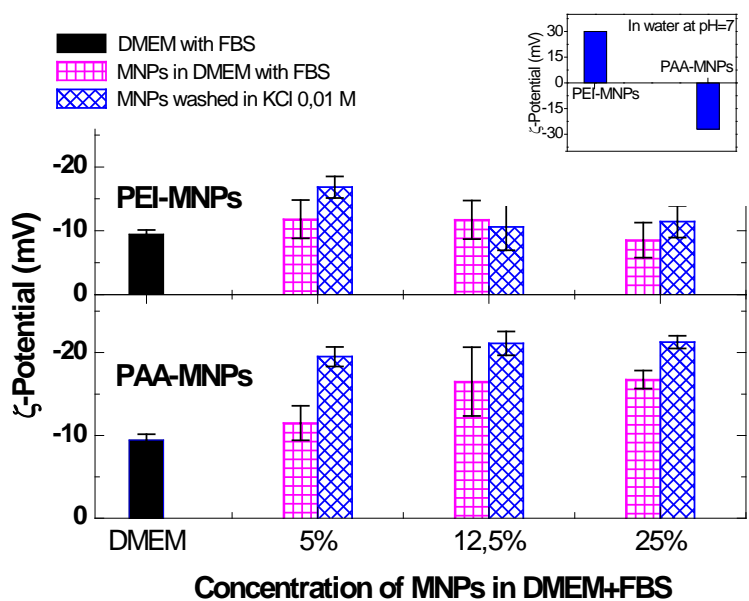

**Figure 6:** *ζ-potential data of PEI-MNPs and PAA-MNPs in cell culture medium (DMEM+15% FBS) before and after washing with water/KCl 0.01M. Different concentrations of MNPs are shown. Note the opposite initial values of the ζ-potential in the as prepared colloids (inset).*

The above results show the effect of the surface chemistry on protein adsorption. As expected, PEI-MNPs having positive zeta potential were found to adsorb more proteins while PAA-MNPs with negative zeta potential showed less protein adsorption. The ζ–potential studies on the DMEM containing FBS indicated mean value of −10 mV (Figure 6). This explains the higher adsorption of the proteins onto the positive PEI-MNPs due to electrostatic interactions. Previous works from Nienhaus et al. [32] have clearly demonstrated the effect of different protein types on the nature of the resulting protein corona in terms of structure, thickness and stability. When compared to the present work, where the variety of proteins present in the culture medium resulted in a non-specific adsorption process, it is clear that a more complex final state should be expected for the MNPs under actual physiological environments.

### *MNP uptake by SH-SY5Y cells: effects of the protein adsorption*

The in vitro experiments were carried out on human neuroblastoma cell line (SH-SY5Y). As a first step we determined the cell viability of SH-SY5Y cell line when incubated with PEI-MNPs and PAA-MNPs for increasing concentrations and incubation times. The analysis performed for both types of MNPs by Trypan blue assays and Flow Cytomtetry after 24, 48 and 72 h of incubation showed only a slight toxicity on this cell line, with no significant differences observed between the PEI- and PAA-coated MNPs (Figure S5 of the supplementary data). Moreover, cell viability levels for both MNPs were similar to the control sample even for the highest amounts added (50 µg/mL of MNPs). In order to obtain trustable information regarding the cell uptake kinetics, we verified the growing rate and doubling time, $t_D$, of the SH-SY5Y cell line under the conditions of our experiments (Figure S4 of supplementary material).

Figure 7 shows the total amount of MNPs uptaken by the SH-SY5Y cells as a function of the total mass of MNPs added, at incubation times of 15 and 72 h. For all concentrations used, the amount of MNPs associated to the cells was much larger for the PEI-MNPs. It is important to mention that a vigorous washing process was performed three times before measurement of MNPs contents was done. Therefore the data of Figure 7 refers to those MNPs either incorporated or strongly attached to the cell membrane (the actual situation will be discussed below). In all concentrations tested, a linear relationship between the total amounts of added and incorporated MNPs was found for both MNPs. At $t = 15$ h (i.e., less than the doubling time $t_D = 16.6$ h) the rate of uptake as a function of concentration could be fitted with a straight line with slopes 0.54(9) and 0.27(2) for PEI- and PAA-MNPs, respectively. These values imply that at $t = 15$ h the cells were able to incorporate only a 54% (PEI-MNPs) and 27% (PAA-MNPs) of the particles available. On the other hand, at $t = 72$ h (that is, $t = 1.67\ t_2$) the increase was also linear, but the slopes 1.03(7) for PEI-MNPs indicated that after replication the new cells were able to incorporate the 100% of the MNPs added, whereas for the PAA-MNPs the slope was 0.58(2), meaning that only 58% of the MNPs could be incorporated.

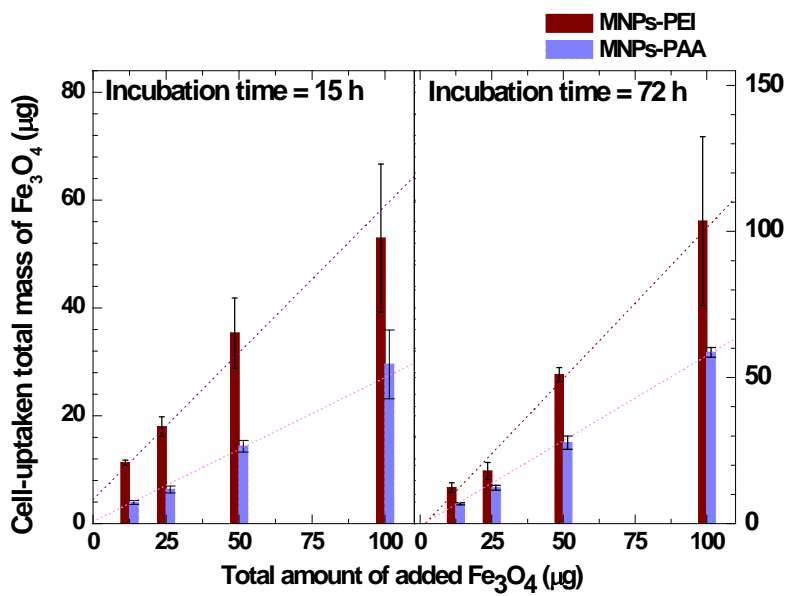

**Figure 7.** *Total cell uptake vs total added amount of PEI-MNPs and PAA-MNPs (at 15 and 72 h incubation time).*

Since doubling of the cell population takes place in any experiment enduring more than the cell doubling time, the actual efficiency for MNPs uptake of the cells must be expressed as the normalized mass of MNPs *per cell* that is incorporated. Accordingly, the uptake kinetics was analyzed from the data of MNPs *per cell* obtained as a function of incubation time in different conditions of MNPs availability (Figure 8). These data corroborated the higher affinity of neuroblastoma cells for PEI-MNPs nanoparticles in all concentration range and incubation times, as compared to the PAA-MNPs uptake. As for the time dependence cell uptake, the behavior was the same for both types of MNPs, with maximum uptake efficiency between 8 and 15 h of incubation.

At the shortest incubation times (30 minutes) the amount of incorporated PAA-MNPs was 2-4 pg/cell irrespective of the added MNP concentration. These values were substantially lower than the concentration dependent values from 5 to 16 pg/cell observed for PEI-MNPs at the same incubation time, indicating that for short times the uptake is small and the attachment to the cell membrane is more important. After 2 h of incubation the differences between PEI- and PAA-MNPs uptake increases abruptly in a concentration-dependent way, reaching a 4.5-fold larger uptake of PEI-MNPs when 100 μg of MNPs were added (Figure 8d).

For incubation times larger than the cell doubling time ($t_D$=16.5 h, see Figure S4 in the supplementary material) a monotonous decrease in the amount of MNPs per cell was observed, as expected for a constant-mass incubation experiment where the MNPs are being split between cells following cell division. However, the possibility of MNPs

being also exocytosed from the cells cannot be ruled out with the present experiments.[33]

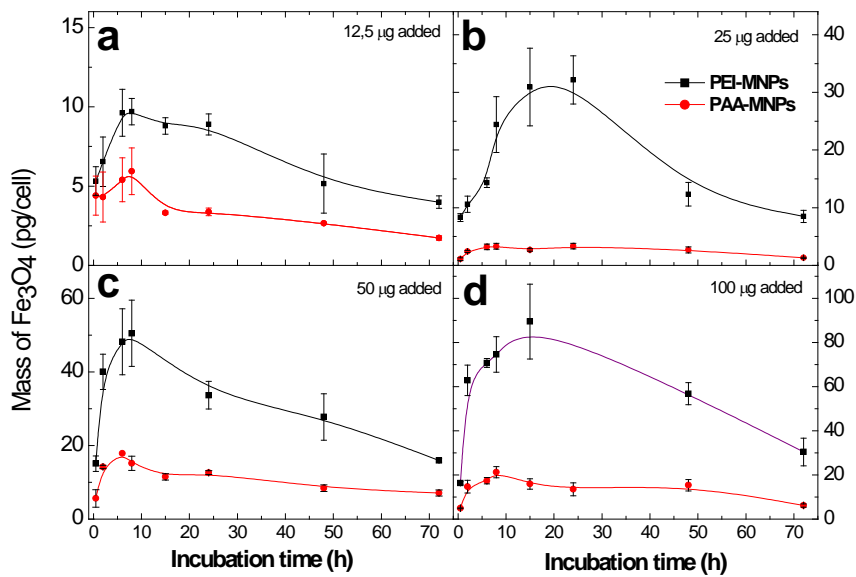

**Figure 8**. *Cellular uptake of PEI-MNPs (squares) and PAA-MNPs (circles) as a function of incubation time and increasing concentration of nanoparticles: a) 12.5 µg; b) 25 µg; c) 50 µg and d) 100 µg of MNPs added. Values represent the mean ± standard deviation of three different experiments. Lines are only a guide to the eye.*

The experimental results displayed here prove that nanoparticles surface chemistry and size determines the cellular binding of nanoparticles. The data reveals that there is a rapid coating of particles by serum proteins and a correlation between protein adsorption to particles surface and cellular binding. The higher adsorption of proteins onto PEI-MNPs seems to favor their uptake by neuroblastoma cells compared to PAA-MNPs. We can assume that the interaction between the nanoparticles and the cells involves the whole MNPs-protein complex and not the bare nanoparticles, and that therefore the properties of this complex are the parameters to influence the uptake by the cells. Indeed, MNPs uptake may be due to a two-step process: NPs covered with protein corona, adherer to the cell membrane and interact with lipid and proteins of the membrane. This step is followed by the activation of some energy-dependent uptake mechanism which allows the NPs to be internalized by the cell.

Proteins on the surface could mediate binding to cells by two mechanisms, specific and non-specific. In specific interactions, particle adsorbed proteins interact with the binding sites of receptor proteins on cell surfaces. The non-specific interactions involve random binding between the proteins on nanoparticles and the components of cell surfaces. Since both particles studied here display similar surface potential ($\approx$ -10 mV) in biological medium, our results seem to support the existence of specific interactions instead of non-specific ones. Multiple serum proteins attached to the nanoparticles may allow entry through multiple receptor sites. It is known that depending on nanoparticle surface charge different proteins are adsorbed.[34] For instance, Gessner et al. observed that positive charged nanoparticles prefer to adsorb proteins with isoelectric point (pI) < 5.5 such as albumin, while the negative surface charge enhances the adsorption of proteins pI > 5.5 such as IgG. [35]. Due to the opposite surface charge of PEI-MNPs and PAA-MNPs, different serum proteins may adhere on nanoparticles surface and therefore influence their uptake. The corona formed onto PAA-MNPs may induce lower adhesion to the cell membrane, affecting their internalization.[36] Thus, not only the amount but also the type of protein adsorbed onto the particles could be important for determining the MNPs uptake efficiency of neuroblatoma cells.

**Cellular localization of PEI-MNPs and PAA-MNPs**

The uptake and intracellular distribution of the magnetic nanoparticles were examined by transmission electron microscopy (TEM) and SEM-FIB Dual Beam techniques. Large amounts of PEI-MNPs were observed inside SHSY5Y cells (Figure 9, central row) whereas PAA-MNPs were found but in lower concentration (Figure 9, lower row). Analysis of the samples by EDS–HAADF spectra confirmed the Fe contents of these clusters, characteristic of the $Fe_3O_4$ nanoparticles. Moreover the high resolution images in both bright and dark field modes also showed that the morphology of the MNPs is preserved inside the cells, ruling out any significant particle degradation (see also figures S6 and S7 of the Supplementary Data). It is important to mention that the analysis of more than 50 cell samples showed that a substantial amount of PEI-MNPs was often present within the cytoplasmic space, whereas for the PAA-MNPs most of the cell slices were empty of MNPs or had few of them in small clusters (Figure S7 supplementary data). Notwithstanding the small number of images in which PAA-MNPs were found inside the cells, for illustrative purposes the last row of Figure 9 contains one of these (statistically not relevant) images including PAA-MNPs. The MNPs seems to be not free in the cytosol, but surrounded by a thin membrane, indicating some endosome-mediated uptake process in both cases.

The fractions of MNPs effectively internalized and those attached to the cell membrane were different for PEI- and PAA-MNPs. The fraction of PEI-MNPs attached onto the cell membrane were found to form clusters of large size (up to ≈500-1000 nm), and the FIB/SEM cross section through these clusters revealed that they crossed the membrane into the cell interior. For the PAA-MNPs, in spite of the much smaller amount of MNPs observed into cells, in all cases the PAA-MNPs were located forming small aggregates within the cytoplasm, with no particles attached to the cell membrane.

As mentioned in the Introduction, several groups have previously observed that negatively-charged MNPs can also be uptaken by cells. Anionic particles are initially adsorbed at specific binding sites (positively-charged) distributed along the cell membrane, which retains negative MNPs thorough electrostatic interaction. This initial interaction is followed by the formation of MNPs aggregates on the cell membrane due to repulsive electrostatic interactions between MNPs and those negatively charged domains of the cell surface. Contrary to the adsorption process dominating the uptake mechanism of cationic MNPs, the internalization capacity for anionic MNPs depends on the cell types involved and is not expected to dominate the overall uptake efficiency.[37, 38]

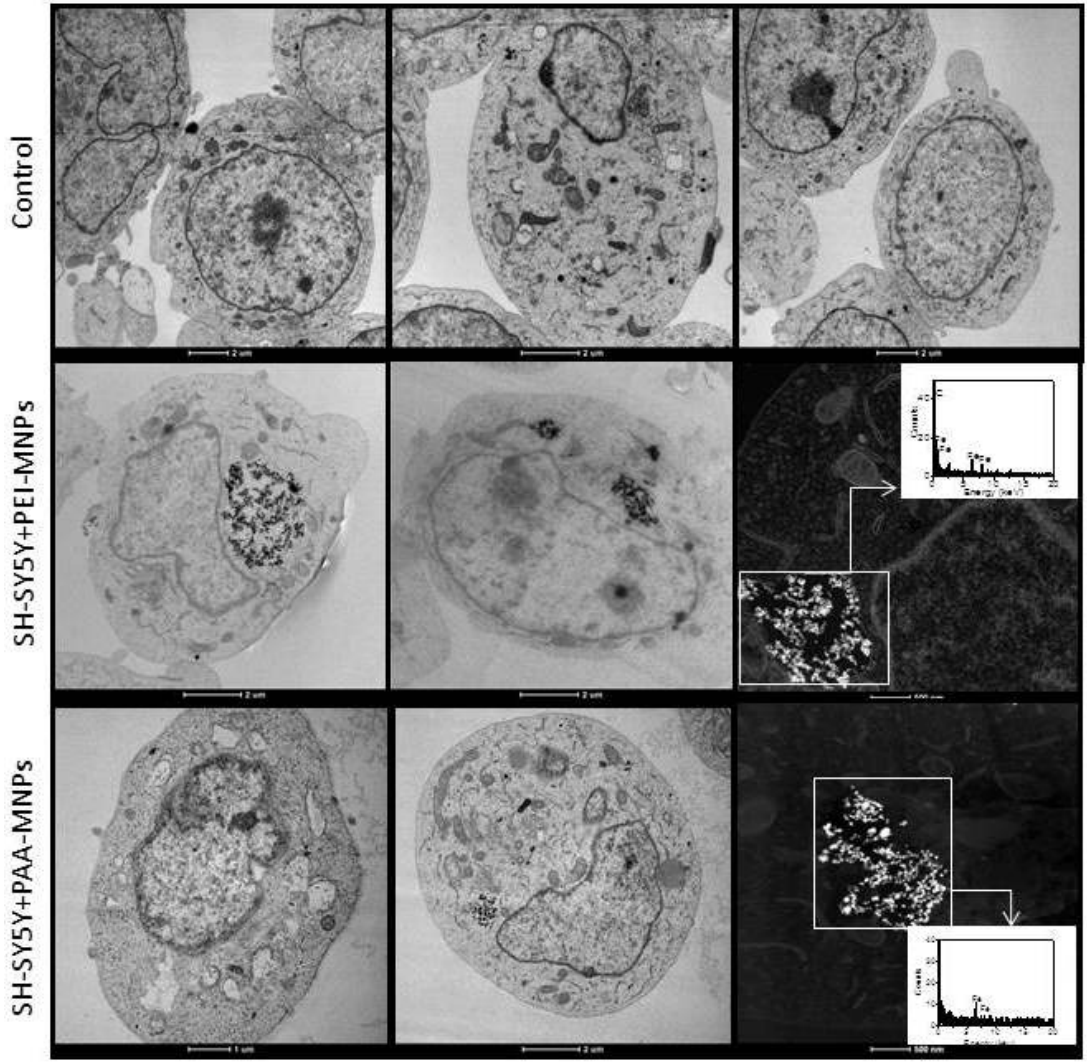

**Figure 9:** TEM and *STEM images of SH-SY5Y control cells (upper row); incubated (24 h; 10 µg/mL) with PEI-MNPs (center row) and PAA-MNPs (lower row); The last column shows the EDS–HAADF spectra of PEI-MNPs (upper) and PAA-MNPs (lower) inside the cell.*

EDX spectra performed on SHSY5Y cells cross-sectioned by FIB/SEM confirmed the particle localization also in the growth cone of the cells (Figure 10).

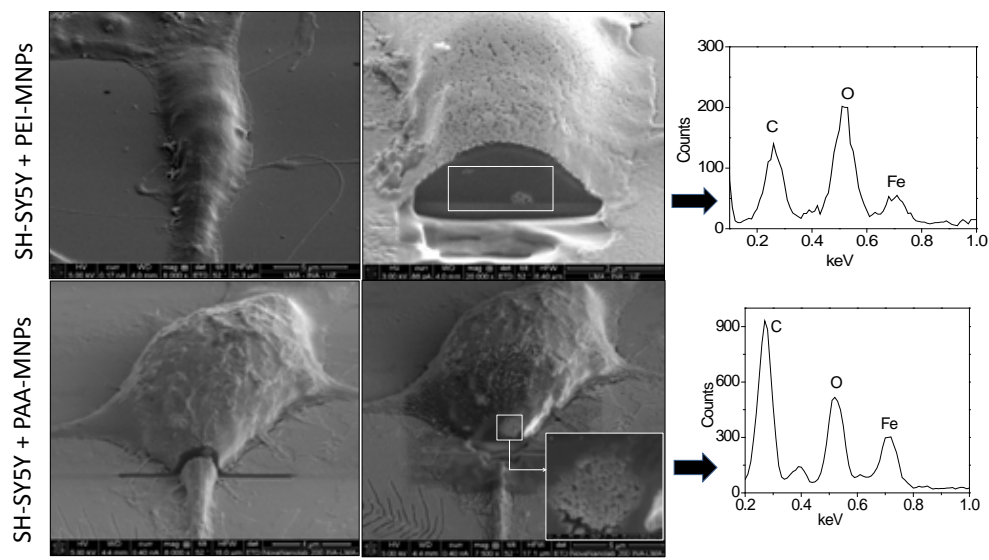

**Figure 10.** *FIB-SEM dual beam analysis of SH-SY5Y cells incubated with 10 ug/l PEI- MNPs* and PAA-MNPs for 24h...

It is known that different types of ligands that bind on the cell membrane of cultured neurons and neuroblastoma cells undergo endocytosis into vesicles and afterwards transferred to the Golgi apparatus.[39] Neuroblastoma cells have a variable number of saturable binding sites for different type of molecules, from 50 to $10^7$ sites/cell. Given the smaller amount of PAA-MNPs uptaken within the first doubling time, we hypothesize that internalization of PAA-MNPs occurs through previous interaction with these binding sites. The TEM and FIB data suggest that, although the average surface charge of the MNPs-protein agglomerates is determined by the proteins, during the uptake a part of the loosely-bounded proteins dissociate from the agglomerates, exposing different interfaces to the cell membrane. It is yet to be determined whether the kinetics of binding to the cell membrane and the incorporation pathways depend on the nature of the proteins involved.

## *Conclusions*

The results reported here illustrated the transformations experimented by colloidal nanoparticles when in contact with biological media, and how they influence the uptake ability of a specific cell line. Using two samples with very similar average size, size distribution and magnetic properties, but opposite charge at the surface, we were able disentangle the influence of surface charge on the formation of the protein-MNPs agglomerates in protein-rich cell culture media. Under in-vitro conditions the time evolution of these protein-MNPs clusters shown by $\zeta$-potential, TGA and dynamic light scattering measurements was found to depend on the free functional groups available at the polymer surface, being bigger for the positively-charged PEI-MNPs. Our results clearly indicate that controlling the non-specific adsorption of proteins to MNPs can be tailored through proper functionalization of their surface

The dynamics of MNPs internalization into SH-SY5Y neuroblastoma cell line was found to depend on the incubation time, with a maximum at 8-10 h of incubation. Although both, PEI and PAA-MNPs could enter the cells, we observed that the mass of internalized/attached PEI-MNPs was much larger than for the PAA-MNPs. While PEI-MNPs were found both strongly attached to the cell membrane and internalized in the form of large clusters, PAA-MNPs were poorly internalized and found to be located almost exclusively into membrane-bound endocytic compartments. The large clusters (up to 700 nm) of PEI-MNPs observed onto the cell membrane remained attached even after vigorous washing the cells several times, indicating a remarkable strength of the binding interaction. We hypothesize that opposite surface charge of PEI- and PAA-MNPs result in adsorption of different proteins that in turn determine different cell internalization pathways. Although the generalization of the above results to other physiological media and to different cell types is yet to be proven, it is clear that a detailed characterization of the MNPs-protein complex must be done to understand the nature of MNPs-cell interactions.

## *Acknowledgments*

The authors are grateful to T.E. Torres for his valuable help on the dual beam analysis of samples. Financial support from the Spanish Ministerio de Economía y

Competitividad (MINECO) (project MAT2010-19326 and HelloKit INNPACTO) is also acknowledged.## *References*

[1] Tenzer S, Docter D, Kuharev J, Musyanovych A, Fetz V, Hecht R, et al. Rapid formation of plasma protein corona critically affects nanoparticle pathophysiology. Nat Nanotechnol. 2013;8:772-U1000.
[2] Peng Q, Zhang S, Yang Q, Zhang T, Wei XQ, Jiang L, et al. Preformed albumin corona, a protective coating for nanoparticles based drug delivery system. Biomaterials. 2013;34:8521-30.
[3] Eberbeck D, Kettering M, Bergemann C, Zirpel P, Hilger I, Trahms L. Quantification of the aggregation of magnetic nanoparticles with different polymeric coatings in cell culture medium. Journal of Physics D-Applied Physics. 2010;43.
[4] Rocker C ZF, Parak WJ, and Nienhaus GU. A quantitative fluorescence study of protein monolayer formation on colloidal nanoparticles. Nature of Nanotechnology. 2009;4.
[5] Darabi Sahneh F, Scoglio C, Riviere J. Dynamics of nanoparticle-protein corona complex formation: analytical results from population balance equations. PloS one. 2013;8:e64690.
[6] Dobay MPD, Alberola AP, Mendoza ER, Raedler JO. Modeling nanoparticle uptake and intracellular distribution using stochastic process algebras. Journal of Nanoparticle Research. 2012;14.
[7] Saito S, Tsugeno M, Koto D, Mori Y, Yoshioka Y, Nohara S, et al. Impact of surface coating and particle size on the uptake of small and ultrasmall superparamagnetic iron oxide nanoparticles by macrophages. Int J Nanomedicine. 2012;7:5415-21.
[8] Lundqvist M. Nanoparticles: Tracking protein corona over time (vol 8, pg 701, 2013). Nat Nanotechnol. 2013;8:806-.
[9] Limbach LK, Li YC, Grass RN, Brunner TJ, Hintermann MA, Muller M, et al. Oxide nanoparticle uptake in human lung fibroblasts: Effects of particle size, agglomeration, and diffusion at low concentrations. Environmental Science & Technology. 2005;39:9370-6.
[10] Safi M, Courtois J, Seigneuret M, Conjeaud H, Berret JF. The effects of aggregation and protein corona on the cellular internalization of iron oxide nanoparticles. Biomaterials. 2011;32:9353-63.
[11] Kralj S, Rojnik M, Romih R, Jagodic M, Kos J, Makovec D. Effect of surface charge on the cellular uptake of fluorescent magnetic nanoparticles. J Nanopart Res. 2012;14:1151.
[12] Jo J, Aoki I, Tabata Y. Design of iron oxide nanoparticles with different sizes and surface charges for simple and efficient labeling of mesenchymal stem cells. J Control Release. 2010;142:465-73.
[13] Villanueva A, Canete M, Roca AG, Calero M, Veintemillas-Verdaguer S, Serna CJ, et al. The influence of surface functionalization on the enhanced internalization of magnetic nanoparticles in cancer cells. Nanotechnology. 2009;20:115103.
[14] Schweiger C, Hartmann R, Zhang F, Parak WJ, Kissel TH, Rivera Gil P. Quantification of the internalization patterns of superparamagnetic iron oxide nanoparticles with opposite charge. J Nanobiotechnology. 2012;10:28.
[15] Kalambur VS, Longmire EK, Bischof JC. Cellular level loading and heating of superparamagnetic iron oxide nanoparticles. Langmuir. 2007;23:12329-36.